\documentclass[twocolumn,showpacs,amsmath,amssymb,pra,floatfix]{revtex4}

\usepackage{graphicx}
\usepackage{dcolumn}
\usepackage{bm}
\usepackage{bbm}
\usepackage{multirow,amssymb,amsbsy,amsmath}
\usepackage{stmaryrd}
\usepackage{color}

\newcommand{\ddim}{\udelta\kern0.1em}

\newcommand{\beikonst}[2]{\left( #1 \right)_{\kern-0.2em #2}}

\newcommand*{\ket}[1]{\mathopen{|}#1\mathclose{\rangle}}

\newcommand{\anticom}[2]{\left\{#1,#2\right\}}

\newcommand{\ketbra}[1]{\mathopen{|}#1\mathclose{\rangle}\hspace{-0.25em}\mathopen{\langle}#1\mathclose{|}}
\newcommand{\ketbrap}[2]{\mathopen{|}#1\mathclose{\rangle}\hspace{-0.25em}\mathopen{\langle}#2\mathclose{|}}

\begin{document}


%
%

\title{Quantum simulation of many-body spin interactions with
  ultracold polar molecules}

\author{Hendrik Weimer}%
\email{hweimer@itp.uni-hannover.de}
\affiliation{Institut f\"ur Theoretische Physik, Leibniz Universit\"at Hannover, Appelstr. 2, 30167 Hannover, Germany}

\date{\today}%

\begin{abstract}

  We present an architecture for the quantum simulation of many-body
  spin interactions based on ultracold polar molecules trapped in
  optical lattices. Our approach employs digital quantum simulation,
  i.e., the dynamics of the simulated system is reproduced by the
  quantum simulator in a stroboscopic pattern, and allows to simulate
  both coherent and dissipative dynamics. We discuss the realization
  of Kitaev's toric code Hamiltonian, a paradigmatic model involving
  four-body interactions, and we analyze the requirements for an
  experimental implementation.

\end{abstract}


\maketitle

\section{Introduction}

The realization of quantum simulators -- devices which can replicate
the dynamics of other quantum systems \cite{Feynman1982,Lloyd1996} --
is currently one of the most exciting topics in the field of ultracold
quantum gases \cite{Bloch2012}. This is particularly relevant in areas
where classical simulation methods have proven to be inadequate due to
the exponential growing Hilbert space dimension, such as in frustrated
quantum magnets, where the system sizes that can be studied using
exact diagonalization methods are typically limited to less than 50
spins \cite{Nakano2011,Lauchli2011}.

Ultracold polar molecules are particularly promising candidates for
the quantum simulation of spin models due to their long coherence
times and strong electric dipole interactions at distances compatible
with optical addressing. In the past, several theoretical proposals
have been made for the realization of quantum magnetism in these
systems
\cite{Micheli2006,Buchler2007a,Gorshkov2011,Gorshkov2011b,Peter2012,Lemeshko2012,Yao2012a,Yao2012b,Gorshkov2013}. However,
these proposals are challenging to extend to higher order many-body
interactions as these terms arise within a perturbation series and
thus become exponentially weaker the more particles participate in the
interaction \cite{Micheli2006,Buchler2007a}. On the other hand, such
spin models with many-body interactions have recently received great
attention in the context of Kitaev's toric code Hamiltonian
\cite{Kitaev2003}, which has interesting topological properties, and
for the generation of cluster states, which are relevant for
measurement-based quantum computing \cite{Raussendorf2005}.

In this article, we describe an architecture for an efficient
non-perturbative simulation of many-body spin interactions. Our
approach is related to a recent quantum simulation proposal based on
strongly interacting Rydberg atoms \cite{Weimer2010,Weimer2011};
however, due to the inherent limitations to Rydberg excitations due to
their fast radiative decay, an implementation with ultracold polar
molecules within their electronic ground state manifold might offer
some beneficial aspects. Our approach to quantum simulation will be a
digital one, i.e., the dynamics of the simulated system arises
stroboscopically at discrete time intervals \cite{Lloyd1996}. In
particular, the effective dynamics will be created by applying
sequences of microwave pulses coupling rotational excitations of the
polar molecules, realizing sequences of single-body and two-body
quantum logic gates. We first describe the setup composed of ultracold
polar molecules trapped in optical lattices, followed by a discussion
on the implementation of a single many-body interaction. Besides
coherent time evolution, the quantum simulator is also capable to
incorporate dissipative dynamics, including efficient preparation of
the ground state. We will then generalize these concepts to the full
lattice system and discuss the sources and consequences of residual
imperfections. Finally, we describe the experimental requirements to
implement our quantum simulation architecture.

\section{Setup of the system}

\begin{figure}[b]

\includegraphics[width=8cm]{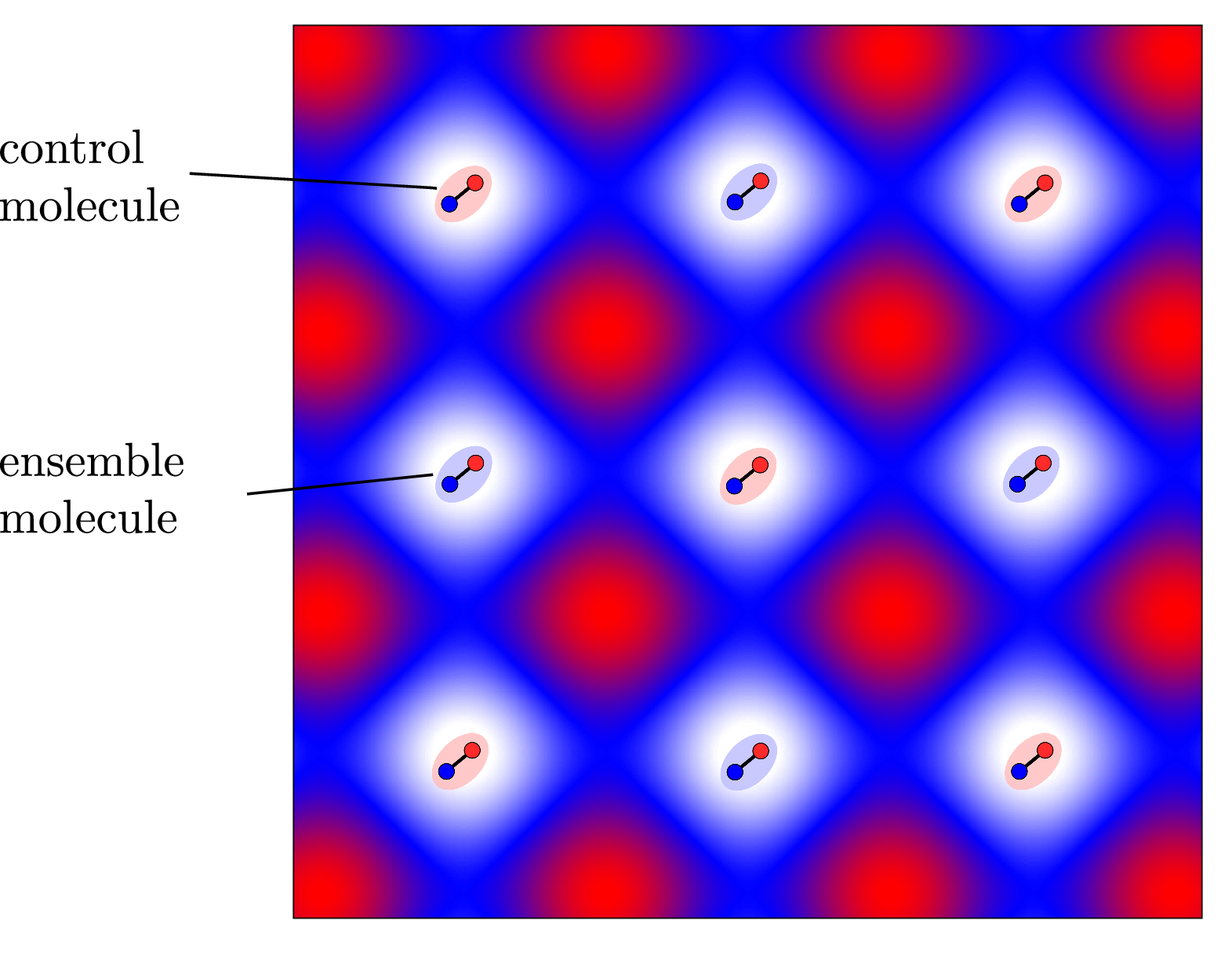}

\caption{Proposed experimental setup. Polar molecules are arranged in
  the intensity minima of a two-dimensional optical lattice with one
  molecule per lattice site. Molecules on neighboring lattice sites
  take different roles as control or ensemble molecules.}
\label{fig:setup}

\end{figure}

We consider a two-dimensional system of ultracold polar molecules in
their rovibrational ground state \cite{Ni2008,Deiglmayr2008} and
loaded into an optical lattice \cite{Miranda2011}, see
Fig.~\ref{fig:setup}. We focus on the case where the lattice potential
is deep enough to suppress any tunneling between different
sites. Furthermore, we assume the molecules to be initially prepared
in a well-defined hyperfine state \cite{Ospelkaus2010a}. Here, we are
also interested in a setup where a single \emph{control molecule} is
surrounded by four \emph{ensemble molecules}, in a way that the former
can be manipulated independently without affecting the other
ones. Such a controllability is most readily achieved if the molecules
can be addressed individually using optical fields
\cite{Weitenberg2011}. Note that the distinction between control and
ensemble molecules is a purely logical one, i.e., both types can be
realized using a single molecular species.

\begin{figure}[t]

  \includegraphics[width=8cm]{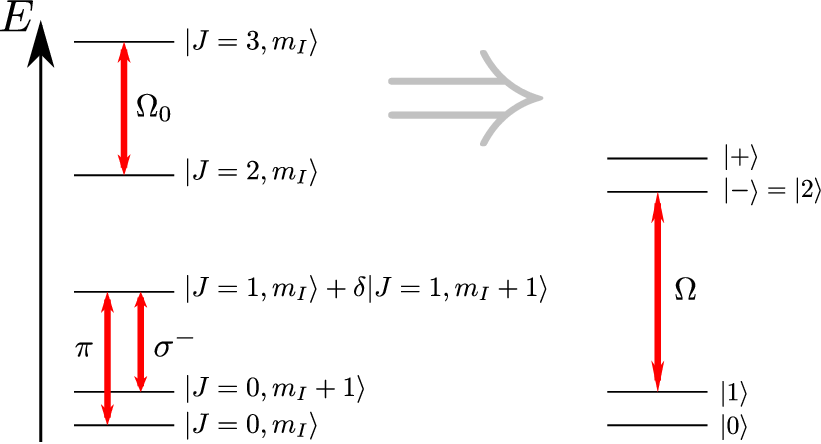}

  \caption{Relevant part of the internal level structure. Depending on
    the polarization of microwave fields ($\pi$,$\sigma^+$),
    hyperfine-preserving or hyperfine changing transitions can be
    driven. A strong microwave field $\Omega_0$ couples the $J=2$
    manifold to the $J=3$ manifold, resulting in dressed states that
    can be accessed by an additional two-photon microwave driving
    $\Omega$. For the purpose of the quantum simulator, only the
    states $\ket{0}$, $\ket{1}$, and $\ket{2}$ are important.}
\label{fig:internal}

\end{figure}

The relevant level structure of the molecules is shown in
Fig.~\ref{fig:internal}, where we focus on the lowest four
rotationally excited states. We consider the driving of rotational
transitions by microwave fields. By combining linearly and circularly
polarized microwaves together with the electric quadrupole interaction
coupling the nuclear spin to the rotation ($\delta$) \cite{Brown2003},
it is possible to selectively drive transitions between hyperfine
states \cite{Ospelkaus2010a}.

Additionally, we are interested in the situation where a strong
linearly polarized microwave field $\Omega_0$ is resonant with the
transition between the states $\ket{J=2,m_I}$ and $\ket{J=3,m_I}$. The
quantization axis is defined by the polarization, so the dynamics will
be constrained to states with vanishing projection of the angular
momentum on the quantization axis, i.e., $J_z=0$.  In the dressed
frame of this driving described by the Hamiltonian
\begin{equation}
H_0 = \Omega_0\ketbrap{J=2,m_I}{J=3,m_I} +\mathrm{h.c.},
\end{equation}
the state $\ket{-} =
(\ket{J=2}-\ket{J=3})/\sqrt{2}$ will pick up an effective permanent
dipole moment of $d_e = 3 d/\sqrt{70}$, where $d$ is the bare electric
dipole moment of the molecule \cite{Lemeshko2012}. Additionally, we
consider a weak microwave driving $\Omega$ of the two-photon
transition between the state $\ket{0}$ and the dressed state
$\ket{-}$.

Finally, the position of the $\ket{-}$ manifold can be shifted
relative to the $\ket{J=0}$ manifold using optical fields resulting in
a differential ac Stark shift \cite{Kotochigova2010}, which can shift
the molecules out of resonance of the microwave field $\Omega$. By
confining the optical potentials to individual lattice sites, single
molecule addressing on a lengthscale of $500\,\mathrm{nm}$ can be
realized \cite{Weitenberg2011}. Further improvements might be achieved
using sub-wavelength addressing techniques using strong field
gradients \cite{Stokes1991} or electromagnetically induced
transparency \cite{Gorshkov2008}.

Essentially, the setup presented here allows to selectively create
rotational excitations, depending of the position of the molecule and
its hyperfine state. Consequently, the dynamics is effectively
constrained to three states: two nuclear states corresponding to the
$\ket{J=0}$ manifold (named $\ket{0}$ and $\ket{1}$ hereafter), and
one state in the $\ket{-}$ manifold ($\ket{2}$). The benefit of this
setup based on microwave dressing compared to ones involving static
electric fields lies in the strong suppression of dipolar flip-flop
terms coupling different $J$ states, thus eliminating a potential
error source for the quantum simulator.

\section{Simulation of the toric code}

\begin{figure}[t]

\includegraphics[width=6cm]{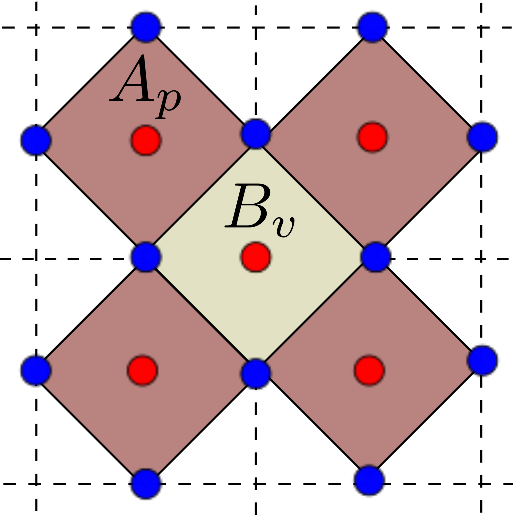}

\caption{Two-dimensional lattice setup for the toric code involving control (red) and ensemble molecules (blue). The ensemble molecules taking part in plaquette operators $A_p$ and vertex operators $B_v$ are colored accordingly.}
\label{fig:toric}

\end{figure}

The specific model we want to outline a quantum simulator for is
Kitaev's toric code \cite{Kitaev2003}. It serves as paradigmatic model
of a large class of so-called \emph{stabilizer Hamiltonians}
\cite{Gottesman1996,Bombin2006}, whose ground states can be found by
local minimization of the energy. The toric code Hamiltonian is given by
\begin{equation}
  H = -E_0 \left(\sum\limits_p A_p + \sum\limits_v B_v\right),
  \label{eq:toric}
\end{equation}
with the ``plaquette'' operators $A_p = \prod_{i\in p}\sigma_x^{(i)}$
and the ``vertex'' operators $B_v = \prod_{i\in v}\sigma_z^{(i)}$
containing Pauli operators representing four-body spin interactions,
see Fig.~\ref{fig:toric}. On a torus, the ground state of the toric
code is fourfold degenerate and corresponds to the $+1$ eigenvalues of
all mutually commuting operators $A_p$ and $B_v$ and exhibits
topological order. Plaquettes or vertices with a $-1$ eigenvalue are
quasiparticles called \emph{magnetic charges} or \emph{electric
  charges}, respectively, and always occur in pairs with a string
operator connecting them \cite{Kitaev2003}. Due to the
non-commutativity of $\sigma_x$ and $\sigma_z$, moving a quasiparticle
of one type around a second one of the other type will result in a
minus sign applied to the quantum state; hence, the quasiparticles of
the toric code are abelian anyons. 

These exotic properties have led to many proposed realizations within
ultracold quantum gases or condensed matter systems
\cite{Duan2003,Micheli2006,Jackeli2009,Weimer2010,Nielsen2010,Weimer2011,Fowler2012},
but despite experimental success in small systems
\cite{Lu2009,Barreiro2011}, a large-scale many-body implementation
revealing the topological properties is still lacking
\cite{Lang2012}. In the following, we will discuss the implementation
of the toric code within our quantum simulator architecture. As all
terms of the Hamiltonian are mutually commuting, the dynamics of the
toric code can be constructed by sequential implementation of the
plaquette terms and vertex terms, i.e.,
\begin{equation}
  U = \exp(-iHt) = \prod\limits_p\exp(iE_0A_pt/\hbar)\prod\limits_v\exp(iE_0B_vt/\hbar).
\end{equation}
The form of the plaquette interactions $A_p$ and the vertex interactions
$B_v$ is identical up to a global rotation interchanging $\sigma_x$
and $\sigma_z$; hence, we focus first on the implementation of a
single $B_v$ term, with all remaining terms of the Hamiltonian to be
realized in an analogous way.

\subsection{Single plaquette interaction}

As a crucial step, we now make use of the control molecules to mediate
the vertex interaction. During this process, we require the
implementation of single qubit rotation between the logical states
$\ket{0}$ and $\ket{1}$. Such single qubit gates between nuclear spin
states have already been experimentally realized based on the electric
quadrupole moment of the nucleus \cite{Ospelkaus2010a}. Furthermore,
our architecture will involve a two-qubit controlled phase gate
between two molecules, which we will present in the following.

\subsubsection{Controlled phase gate}

The two logical states $\ket{0}$ and $\ket{1}$ do not exhibit any
dipole-dipole interactions, hence, it is necessary to first transfer
the population from the $\ket{1}$ state to the $\ket{2}$ state by a
two-photon microwave pulse. Doing this simultaneously in the control
molecule and a single ensemble molecule requires the microwave driving
$\Omega$ to be much stronger than the dipole-dipole interaction,
\begin{equation}
V_{dd} = \frac{1}{4\pi\varepsilon_0}\frac{d_e^2}{a^3},
\label{eq:vdd}
\end{equation}
where $a$ is the separation between the control and the ensemble
molecule. After the pulse, the molecules experience the dipole-dipole
interaction for the time $t_\pi = \pi\hbar/V_{dd}$, following a second
microwave pulse transfering the population back to the rovibrational
ground state. Then, up to local rotations which can be canceled by
appropriate additional driving fields, the system has realized a
conditional phase gate of the form
\begin{equation}
U_{CP}^{(e)} = \ketbra{0}^{(c)} \otimes 1^{(e)} + \ketbra{1}^{(c)} \sigma_z^{(e)},
\end{equation}
where $c$ and $e$ refer to the control and ensemble molecule,
respectively.

\subsubsection{Digital simulation procedure}

We will now turn to the quantum simulation of the vertex term $B_v$ in
the toric code Hamiltonian (\ref{eq:toric}).  During a single timestep
$\tau$ of our digital quantum simulator, a single vertex undergoes the
time evolution $U = \exp(iE_0B_v\tau)$. Such a dynamics is realized by
a threefold sequence, provided the control molecule is initially in
$\ket{0}_c$ \cite{Weimer2010,Weimer2011}: (i) A sequence of quantum
gates maps the eigenvalue $\pm 1$ of the operator $B_v$ acting on the
ensemble molecules is mapped onto the states $\ket{0}$ and $\ket{1}$
of the control molecule, i.e., performing the operation
\begin{eqnarray}
U_{\mathrm{MAP}} &=& \ketbra{0}_c\ketbra{B_v=1}\nonumber\\ &+&
(\ketbrap{0}{1}_c+\ketbrap{1}{0}_c)\ketbra{B_v=-1}.
\end{eqnarray}
 (ii) A single qubit
rotation of the form $U_Z(\phi)=\exp(i\phi\sigma_z^{(c)})$ is applied to the
control molecule. (iii) The mapping of step (i) is undone by the
inverse gate sequence. The total gate sequence for the simulation of
the time evolution reads
\begin{eqnarray}
  U &=& \exp(i\phi B_v)\\ &=& U_{\pi/2}^{(c)} \prod\limits_e U_{CP}^{(e)} U_{-\pi/2}^{(c)} U_Z(\phi) U_{\pi/2}^{(c)} \prod\limits_e U_{CP}^{(e)} U_{-\pi/2}^{(c)},\nonumber
\end{eqnarray}
where $U_{\pi/2}^{(c)} = \exp(i\sigma_y \pi/4)$ is a $\pi/2$ rotation
of the control molecule. This sequence involves a series of four
controlled-phase gates $U_{CP}$ acting on the control molecule and
each ensemble molecule sequentially, which effectively represents a
multi-qubit gate, where the control molecule conditionally manipulates
the ensemble molecules, sometimes denoted as a Controlled-Phase$^N$
quantum gate. This sequential operation offers maximum speed of the
gate sequence as all operations can be carried out resonantly. On the
other hand, conceptually slightly simpler but slower gate sequences
based on direct multi-qubit gates \cite{Muller2009} also exist
\cite{Weimer2010,Weimer2011}.

The effective energy scale $E_0$ of the simulated Hamiltonian depends
on the phase $\phi$ written onto the control molecule during each
timestep $\tau$, i.e., $E_0=\hbar\phi/\tau$. While for the toric code
involving only mutually commuting operators, $\phi$ can be arbitrarily
large, models with non-commuting degrees of freedom mandate the
introduction of a Trotter expansion of the form $\exp(-iH\tau) =
\prod_i \exp(-iH_i\tau) + O(\tau^2)$, requiring $\phi \ll 1$ to
reproduce the desired dynamics with high accuracy.

\subsubsection{Dissipative state preparation}

As already mentioned, the ground state of the toric code Hamiltonian
(\ref{eq:toric}) can be found by locally minimizing the energy of the
stabilizer operators $A_p$ and $B_v$. Therefore, it is possible to
engineer the dynamics such that energy is constantly removed from the
system and the ground state arises as the final state of the dynamics
\cite{Aguado2008,Verstraete2009,Weimer2010,Weimer2011}. As a crucial
element, the dissipation of energy requires the presence of an
incoherent process in the dynamics; here, we are interested in the
situation where the control molecule can be incoherently pumped from
the state $\ket{1}$ to the state $\ket{0}$. However, such a
dissipative step is not as straightforward to realize as with atoms
since the radiative decay of molecules to many vibrational states
makes optical pumping very challenging. Essentially, possible
experimental realzations of the desired dissipative dynamics can be
generalized into three distinct classes: (i) Direct laser cooling of
molecules \cite{Shuman2010,Manai2012}, giving rise to sufficiently
high optical pumping efficiencies. (ii) Microwave driving of the
$\ket{1}$ state to a rotationally excited state, which is strongly
coupled to the $\ket{0}$ state via a lossy microwave stripline
resonator \cite{Andre2006,Wallquist2008}. (iii) Employing
hyperfine-preserving STIRAP processes to dissociate molecules back
into atoms, followed by optical pumping of the atoms and reversing the
STIRAP process.

For the implementation of dissipative state preparation, we employ the
gate sequence
\begin{equation}
  U =  U_{CNOT,i}U_{\pi/2}^{(c)}U_{CP}U_{\pi/2}^{(c)},
\end{equation}
where $U_{CNOT,i}$ is a controlled-NOT gate with the control molecule
as a control qubit and the $i$th ensemble molecule acting as a target
qubit. This two-body gate can be easily constructed from the
controlled phase gate $U_{CP}$ using additional single qubit
rotations. The action of the gate sequence $U$ can be understood by
looking at the states $\ket{\pm,\lambda}$ having the property $\langle
B_v \rangle = \pm 1$ (here, $\lambda$ encodes the information about
the state that is independent of $B_v$). The chosen gate sequence will
yield
\begin{equation}
  U\ket{0}_c\ket{\pm,\lambda}_e = \frac{1\pm 1}{2} \ket{0}_c\ket{+1,\lambda}_e+\frac{1 \mp 1}{2}\ket{1}_c\ket{+1,\lambda}_e.
\end{equation}
Note that both summands contain the state $\ket{+1,\lambda}_e$
for the ensemble molecules, as the final controlled-NOT gate
transforms the state $\ket{1}_c\ket{-1,\lambda}$ into
$\ket{1}_c\ket{+1,\lambda}$ while leaving the
$\ket{0}_c\ket{+1,\lambda}$ state untouched.  
Performing the
dissipative step $\ket{1}_c \to \ket{0}_c$ results in the state
$\ket{+1,\lambda}$ for the ensemble molecules, independent of
$\lambda$. In general, the dynamics created by $U$ and subsequent
dissipation can be described in terms of a discrete time quantum
master equation in Lindblad form,
\begin{equation}
  \rho(t+\tau) = c\rho(t)c^\dagger - \frac{1}{2}\anticom{c^\dagger c}{\rho(t)},
  \label{eq:master}
\end{equation}
with the quantum jump operator $c =
i\sigma_x^{(i)}(1-B_v)/2$. This jump operator consists of the
projector $(1-B_v)/2$ forming an \emph{interrogation part} that checks
whether the system is in a $+1$ eigenstate of $B_v$ and a \emph{pump
  part} $\sigma_x^{(i)}$, which flips a single spin and thus turns any
state with $B_v = -1$ into a state with $B_v = +1$. From the toric
code Hamiltonian it is obvious that the jump operator changing
$B_v=-1$ to $+1$ will lower the energy of the system and thus
constitute a cooling effect. Indeed, by performing this cooling
operation many times on all vertices and plaquettes, we can prepare
the ground state of the toric code \cite{Weimer2010,Weimer2011}.

\subsection{Full lattice model}

\begin{figure}[b]

\includegraphics[width=8cm]{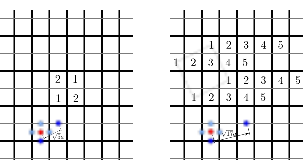}

\caption{Sublattice schemes for the toric code. Numbers from 1 to $z$
  indicate which plaquettes are being addressed simultaneously. As
  exemplified for a single control atom (red) performing a two-qubit
  gate with the ensemble molecule located immediately below (dark
  blue), there is crosstalk with an ensemble molecue associated to a
  different plaquette. Inactive ensemble molecules surrounding the
  control molecule that are are shown in light blue for reference. For
  a simple $z=2$ partitioning scheme, the dominant contribution to the
  crosstalk occurs at a distance of $\sqrt{5}a$, while for a $z=5$
  scheme with a knight's move unit cell, the distance is increased to
  $\sqrt{17}a$.}
\label{fig:part}
\end{figure}

As shown above, it is possible to construct a quantum simulation of
the toric code for both coherent and dissipative dynamics by iterating
over all plaquettes and vertices in the full lattice system. For
maximum performance, it is desirable to exploit the largest degree of
parallelism, while keeping errors at a minimum. While it is possible
to address many molecules in parallel at the same time, one has to be
aware of a few limitations. First, one has to make sure that any
parallel implementation of plaquette and vertex operators does not try
to address the same molecule twice during a single operation. This is
naturally enforced by operating plaquette and vertex operations
sequentially and partitioning the system into two different
sublattices ($z=2$), see Fig.~\ref{fig:part}. Second, the long-range
tail of the dipole-dipole interaction will introduce errors due to
crosstalk between molecules on different plaquettes or vertices. This
effect can be reduced by increasing the unit cell of the
sublattice. Already by using a $z=5$ sublattice arranged in a knight's
move pattern, the crosstalk can be reduced to a percent level
perturbation.

\subsection{Imperfections}

In any realistic experimental situation, there will be other sources
of imperfections besides crosstalk between distant molecules due to
the long range dipolar interaction, such as phase noise of the laser
fields used for optical addressing of individual molecules. Here, the
dominant source of error will be related to the two qubit controlled
phase gates. Effectively, such errors can be described in terms of a
random phase flip $\sigma_z$ acting on one of the qubits with
probability $\varepsilon$. These gate errors will lead to additional
noise terms in the quantum master equation \cite{Weimer2010}, and for
the toric code will result in a finite anyon density $n$, which can be
cast as an effective temperature $T$, according to
\begin{equation}
  T \approx - \frac{2E_{0}}{k_B \log n}.
\end{equation}

%

\begin{figure}[t]

\includegraphics{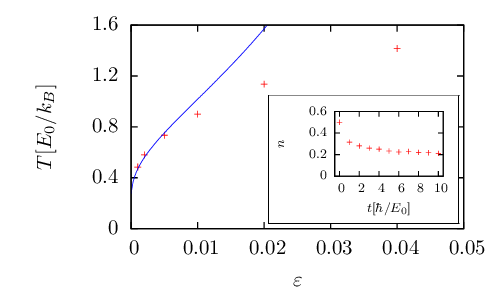}

  \caption{Dependence of the residual temperature $T$ on the two qubit
    gate error probability $\varepsilon$. Numerical simulation for a
    system of 16 ensemble molecules averaged over 1000 realizations are
    shown as crosses, while the solid line is derived within linear
    response theory. The inset shows the relaxation dynamics for
    $\varepsilon = 0.01$.}
\label{fig:etoric}
\end{figure}

Applying linear response theory to the quantum master equation
(\ref{eq:master}), we can obtain an asymptotic expression for the
anyon density $n = 4c\varepsilon$, with the numerical constant $c$
depending on the details of the error model. In the case of phase
errors, we find $c=7/2$. As shown in Fig.~\ref{fig:etoric}, this
behavior is reproduced by numerical simulations in the limit of small
gate errors. Remarkably, already for gate errors on the order of
$\varepsilon \approx 0.01$, the stationary state reaches a low
temperature regime characterized by $k_BT/E_0 < 1$, where
topological order can be detected within finite-size systems
\cite{Castelnovo2007,Nussinov2008}.

\section{Experimental requirements}

To be specific, we focus on an implementation using NaK molecules
\cite{Wu2012}, which combine chemical stability \cite{Zuchowski2010}
with a relatively large electric dipole moment of
$d=2.7\,\mathrm{D}$. Single qubit operations can be carried out with a
speed of $\sim 20\,\mathrm{kHz}$ \cite{Ospelkaus2010a} and hence will
be much faster than the two-qubit conditional phase gates. For the
latter, we obtain a gate speed of $t_\pi^{-1} \approx 2\,\mathrm{kHz}$
for an $a=532\,\mathrm{nm}$ optical lattice, according to
Eq.~(\ref{eq:vdd}). The overall timescale $E_0$ for the simulation of
the toric code will be lower due to the subsequent implementation of
plaquette and vertex operators and the need for partitioning the
system into several sublattices. Here, for the aforementioned $z=5$
partition scheme, we obtain $E_0 = h\times 40\,\mathrm{Hz}$, which is
compatible with typical timescales in experiments with ultracold polar
molecules. For comparison, although the Rydberg quantum simulator can
reach energy scales up to three orders of magnitudes larger
\cite{Weimer2010}, the polar molecule quantum simulator can achieve
similar results for coherence times on the order of $t_c\sim
100\,\mathrm{ms}$, corresponding to a two qubit gate error of
$\varepsilon \approx t_\pi/t_c = 0.005$.

\section{Conclusion and outlook}

In summary, we have demonstrated the feasibility of a quantum
simulator for many-body spin interactions using ultracold polar
molecules trapped in optical lattices. The proposed architecture
includes both coherent and dissipative dynamics and is quite robust
against experimental imperfections, as we have exemplified for the
toric code Hamiltonian. At the same time, the setup naturally allows
for extensions to even larger classes of strongly correlated spin
models, paving the way towards the realization of a universal quantum
simulator based on ultracold polar molecules.

\begin{acknowledgments}

We acknowledge fruitful discussions with S.~Ospelkaus.

\end{acknowledgments}



\end{document}